\documentstyle[12pt]{article}

\begin{document}
\begin{titlepage}
\begin{center}

April 15, 2004     \hfill    LBNL-54884 \\

\vskip .5in

{\large \bf Comments on Shimony's Analysis}
\footnote{This work is supported in part by the Director, Office of Science, 
Office of High Energy and Nuclear Physics, Division of High Energy Physics, 
of the U.S. Department of Energy under Contract DE-AC03-76SF00098}

\vskip .50in
Henry P. Stapp\\
{\em Lawrence Berkeley National Laboratory\\
      University of California\\
    Berkeley, California 94720}
\end{center}

\vskip .5in

\begin{abstract}
Shimony's method of analysis does not distinguish adequately between a legitimate 
assumption of no faster-than-light action in one direction and the to-be-proved 
assertion of faster-than-light tranfer of information in the opposite direction. 
The virtue is noted of replacing the logical framework based counterfactual 
concepts by one based the concept of fixed past and open future.

\end{abstract}

\end{titlepage}

\newpage
\renewcommand{\thepage}{\arabic{page}}
\setcounter{page}{1}

Professor Shimony's article[1] is an extremely helpful contribution to the subject. It summarizes 
in a lucid way the large areas of agreement between us, and provides a back-to-basics proof of the 
two propositions that are the main technical results of my paper[2]. Shimony's long and detailed 
derivation of those two basic propositions should lay completely to rest all but one of the objections 
that were raised against my more compact 1997 proof[3]. I shall examine presently that remaining 
objection, but first will emphasize some key points of agreement mentioned by Shimony.
  
Shimony identifies the motivation of my work, namely the fact that the theorems of J.S. Bell[4] and 
his followers[5] rest explicitly or implicitly on the local-hidden-variable assumption that the values 
of the pertinent observables exist whether they are measured or not. That assumption conflicts with 
orthodox quantum philosophy, and that fact undermines the idea that some sort of faster-than-light 
transfer of information is implied by the conjunction of Bell's theorem and the assumed validity of 
the predictions of quantum theory. The more likely conclusion, from the orthodox perspective, is a 
failure of the hidden-variable assumption. The orthodox interpretation of Bell's theorem is not that 
faster-than-light transfer of information exists. It is rather that the hidden-variable assumption is 
false. Shimony notes that a proof not requiring a hidden-variable assumption of the need in quantum 
theory for faster-than-light information transfer ``would be a profound scientific and philosophical 
achievement.''

Shimony questions the sufficiency of my reasons for supplementing my 1997 proof with the 2004 version[2]. 
He examines, consequently, not my new proof but rather the explicitly counterfactual approach that I 
proposed in a published reply to his earlier comments. That approach differs fundamentally from the 
one used in my 2004 paper, but his proof of the validity of the two propositions covers both formulations.

The proof constructed and criticized by Shimony lies within the general framework of counterfactual 
reasoning, whereas my 2004 proof, although retaining some of the trappings and language of counterfactual 
argumentation, is based on a substantially different foundation. The combination of my assumptions of 
``free choices'' and of ``no backward-in-time influence'' amounts to the assumption that theories 
covered by my new work are to be compatible with the idea of ``fixed past, open future''.  This 
conceptualization circumvents, at the foundational level, the need for counterfactuals. It accords 
with the notion of an advancing ``now'' in which events occur that ``fix and settle'' first the 
free choice made by any agent about which experiment he will perform, and later the outcome of that 
freely chosen experiment. The future is ``open'' in the sense that the choices in regions R and L 
of which experiments are to be performed in those regions are required to be treatable, within 
the theory, as free choices that are made by the agents when the moment ``now'' arrives. The subsequent 
``choice of the outcome of the freely chosen experiment'' is likewise required to be treatable, within 
the class of theories to which the propositions apply, as undetermined until the advancing moment 
``now'' arrives, at which time the outcome also becomes ``fixed and settled''.  These latter choices 
are termed ``nature's choices'' and are required to conform to the statistical rules of quantum theory.  
Treating the theory in this way is supposed to be one adequate way of expressing the content of the 
theory, although not necessarily the only possible way. 

This switch from an approach formulated in the framework of ``counterfactuals'' to one formulated 
in the framework of ``fixed past, open future'' has no significant effect on the proofs of the two 
propositions. But it brings the concepts being used into closer accord with those of orthodox quantum 
thinking. Although philosophers contend that counterfactual concepts pervade science, and are needed 
for science, the significance of results based on the use of counterfactuals remains somewhat shakey 
in the minds of most quantum physicists. But the idea that the events already observed in the past 
by somebody can be treated as if they are fixed and settled, and that our future choices can be 
treated as if they free, agrees with the way that physicists deal with their theories, with their 
theoretical practices, and with their lives in general. 

Shimony's objection to my interpretation begins with the assertion ``But SR is not an assertion 
about actually occurring events. It is a counterfactual conditional''.  This statement alone activates 
the intuitive distrust of scientist in arguments based on counterfactuals. I shall deal presently with 
Shimony's specific objection, raised within the framework of the counterfactual formulation. But first 
I shall describe the application of the two propositions from the ``fixed past, open future'' point 
of view that is more congenial with the normal thinking of quantum physicists.

Why does Shimony claim that the validity of these two propositions lacks scientific significance? 

This wording is not exactly the way that Shimony put it. But scientific significance is the basic 
issue. The theorems of Bell and his followers are ultimately of value because they rule out certain 
possible models or theories of nature. The pertinent questions are thus: Does the joint validity of 
my two propositions rule out some models or theories of nature that are not ruled out by Bell's 
theorems? And does the joint validity of these two propositions rule out all of the local-hidden 
variable theories that are eliminated by Bell's theorem?  If the joint validity of these two 
propositions does indeed rule out all of the hidden-variable theories covered by Bell's theorem, 
and other theories besides, then these propositions are jointly stronger than Bell's Theorem, both 
because their consequences are stronger, they rule out more theories, and also because their 
assumptions are weaker. In this connection it is important to notice that it is not nature that 
is required to conform to the assumptions. It is rather that a theory must, in order for these 
propositions to be applicable to that theory, be such that the choice made by the experimenter 
in the later region R can be treated as a free variable, effectively undetermined until the 
moment of the decision, and that whatever outcome has already been observed in the earlier region L 
can be considered to remain undisturbed by the subsequent events. The premises of the two 
propositions are thus conditions on the class of theories to which these propositions apply.

To see how this works, suppose you are trying to construct a local theory that agrees with the 
predictions of Quantum Theory. Then what has been proved is that if this theory is merely such 
that (1) the experimenter's choices can be considered ``free'' (i.e., without any relevant causal 
roots), and (2) what is observed to happen in region L can be considered to be fixed and settled 
independently of whether R1 or R2 will later be freely chosen and performed by the experimenter 
in region R, and (3) the predictions of quantum theory for the Hardy experiments are valid,  
then the theory must, for these experiments, satisfy the following two properties:

I. If L2 is performed in L, then if R2 were to be performed and were to give outcome + then 
if R1 were to be performed, the outcome would always be $-$.

II. If L1 is performed in L, then if R2 were to be performed and were to give outcome + then 
if R1 were to be performed the outcome would sometimes be $+$. 	
 
I have eliminated here the counterfactual terminology that was employed in my counterfactual-based 
1997 paper, and that was retained in my 2004 paper for the sake of historical continuity. I have 
adopted here the language appropriate to the assumptions of my 2004 paper, which, as emphasized above, 
are concordant with the idea of fixed past open future. The two propositions pertain to the structure 
of a theory in which the free variables are the open choices to be made by the experimenters as to 
which experiments will be performed, and the two propositions are assertions pertaining to relationships 
that then follow from the combination of the assumption of the validity of the relevant predictions of 
quantum theory, together with the idea that the outcomes that have already been observed by some human 
witness in one region can be treated as fixed and settled, and hence completely unalterable by 
subsequent free choices made in a region space-like separated from the first.    

Although something akin to hidden variables might be entailed by these propositions, any such 
structure is here a consequence of our fixed past, open future assumptions, together with the 
assumptions of the validity of predictions of QM. These consequences are not as strong as Bell's
hidden-variable assumptions. 
  
These two propositions, taken together, entail the presence in region R of information about 
the free choice made in L between L1 and L2: no theory that satisfies these to propositions 
can be ``local'' in the sense that it is logically compatible with an exclusion of all 
faster-than-light transfers of information.

No local-hidden-variable theory can satisfy both of these properties: In such a theory the 
observable properties are fixed and definite whether they are measured of not, and they do not 
depend upon which experiment is chosen and performed far away. That combination of conditions 
is not compatible with the validity of these two propositions.   

No Bell-type hidden-variable assumption has entered into the proof of the two properties. These 
two propositions are consequences simply of the assumptions that the theory is compatible with 
the theoretical concept of ``fixed past, open future'', in conjunction with the validity of 
predictions of quantum theory for this Hardy-type experiment. 

Relativistic quantum field theory (RQFT) is compatible with the premises of the propositions. 
This is shown by the works of Tomonaga[6] and Schwinger[7] (TS), where an advancing surface 
``now'' is a parameterized space-like surface $\sigma(\tau)$ such that for $\tau'$ less than 
$\tau$ the surface $\sigma(\tau')$ lies nowhere later than $\sigma(\tau)$, but is 
somewhere earlier. In the TS formulation there exists a fixed history of the evolution of 
the state vector $\Psi(\sigma(\tau))$ for all $\sigma(\tau)$ up until the present time ``Now''. 
In that formulation there is also, in association with the fixing of any outcome, a change 
of the state vector $\Psi(\sigma(\tau))$ that produces an instantaneous transfer of information 
along the space-like surfaces $\sigma(\tau)$.  But in spite of the existence within the TS 
formulation of RQFT of this instantaneous information transfer along space-like surfaces, 
all the predictions of the theory about outcomes of measurements conform to the requirement 
of relativity theory that no such prediction pertaining to an experiment performed in one 
space-time region can depend upon which experiment is chosen and performed in a second 
space-time region that is situated space-like relative to the first. 

The fact that the TS formulation of RQFT does involve faster-than-light information transfers 
does not by itself entail that the existence of such transfers is an intrinsic feature of RQFT 
itself. There are other formulations that focus directly on connections between observables, 
and in which no trace of faster-than-light information transfer is evident. However, application 
of the two propositions requires merely that theory under examination be ``compatible with''
the concepts of ``fixed past, open future'', in the sense that, without altering the content 
of the theory, the choice between R1 and R2 can be treated as free, and the outcome of the 
earlier observation in L can be treated as fixed and settled prior to the fixing of the later 
choice in R  The validity of this assumption is entailed by the TS formulation of RQFT, 
and hence RQFT itself is covered by arguments based on the validity of the two propositions. 

The general conclusion is that no theory that can be treated in accordance with the idea of 
fixed past, open future, and that accords with the quantum predictions for the Hardy experiments, 
can be reconciled with a locality requirement that bans all faster-than-light transfer of 
information. The requirement of compatibility with the idea of fixed-past-open-future is 
compatible with orthodox quantum thinking and is weaker than Bell's assumption of hidden 
variables, which contradicts orthodox quantum thinking. 

We now turn to the two key questions: 

1. Does Shimony's analysis expose any flaw in this fixed-past-open-future argument?

2. Does Shimony's analysis expose any flaw in the corresponding counterfactual-based argument?
 																					
I shall argue that the answer to both questions is No!

Shimony asserts that ``The error in Stapp's argument is his claim that SR is a statement 
about region R alone.'' But what I actually said, as he correctly recorded, was that 
``the truth or falsity of SR is defined by conditions on the truth or falsity of statements 
describing possible events located in region.''. This difference in wording is significant. 
My argument, given above, is based on my wording: I proved two propositions that both follow 
from the stated assumptions, but that are---because of the fact that their truth or falsity 
is defined by conditions on the truth or falsity of statements describing events located 
in R--jointly incompatible with a ban on transfer of information to R of the choice between 
L1 and L2 made by the experimenter in region L. Shimony treats the entire statement SR, 
which involves counterfactuals, as a unit that incorporates, within itself, my key assumption 
that what happened in L was fixed and settled before the decision between R1 and R2 was made, 
whereas I take this key assumption to be a restriction on the class of theories within which 
the pair of propositions is proved to be true. 

This latter approach of taking the stated assumptions to be conditions on the class of 
theories in which the two propositions are jointly true is a direct and completely legitimate 
way to proceed. Incorporating the key assumption of no-backward-in-time influence into the 
meaning of a counterfactual statement is less satisfactory for two reasons. In the first place 
the mere use of statements about events that in principle can never happen, because some 
contrary thing has been asserted to have definitely happened tends by itself to render the 
argument less than ideally rock solid in the minds of physicists. On the other hand, speaking 
directly about properties of a class of theories that satisfy certain specified conditions that 
are themselves in line with quantum philosophy, and are actually satisfied by relativistic 
quantum field theory, is a far more transparent approach that is less likely to enshroud 
subtle difficulties. The second reason is the closely connected fact that the scrambling 
the key causality assumption into the meaning of the words that express contrary-to-fact 
assertions opens the door to possible confusion.
 
The essential point here is that one must be careful not to introduce any assumption that 
injects implicitly into the theory the transfer of information from L to R that the joint 
validity of the two propositions reveals to be present. Shimony's criticism possesses a 
certain initial aura of credibility due to the fact that introducing any causal connection 
between events in R and in L harbors the danger of injecting implicitly some hidden assumption 
of the very influence from L to R that the argument eventually reveals. If a hidden assumption 
of an influence from R to L is smuggled into the assumptions then the fact that such a connection 
eventually emerges would lack significance. On the other hand, if no such assumption is smuggled 
in, and the conclusion that there must be transfer of information from L to R follows logically 
from completely legitimate assumptions that include in an essential way the pertinent predictions 
of quantum theory, then the conclusion pertaining to the theories in question must be deemed to 
be logically valid.

It is well-know that quantum theory is completely compatible with the absence of faster-than-light 
influences in one direction, provided such influences are allowed in other directions. The question 
at issue is whether one can simultaneously forbid faster-than-light influences in all directions. 
Hence if we wish to prove the need for faster-than-light influence in some direction then we can 
legitimately proceed by excluding faster-than-light action in one direction, say right to left, 
and then showing that this restriction, when combined with the assumption of the validity of 
pertinent predictions of quantum theory, entails the need for faster-than-light transfer of 
information in the other direction, namely from left to right. This is the completely legitimate 
line of argument that I employ.

The first part of this legitimate argument is implemented by my assumption that the earlier 
observed outcome in L is fixed and settled, independently of what the later free choice in 
R will be. This assumption of not a hidden assumption of the existence of an action from left to right. 
It is the completely-legitimate-in-this-context demand that there be no action from right to left. 
This assumption, by itself, does not entail any influence from left to right. Only when combined 
with the predictions of quantum theory does it lead to the conclusion that there must be 
information transfer from left to right.  Thus the requirement of no action from right to left, 
whether regarded as a condition on the class of covered theories, or as part of the meaning of SR, 
is completely legitimate, in the context of this proof. But Shimony's analysis does not distinguish 
this completely-legitimate-in-this-context assumption of no action from right to left from what 
would be a completely illegitimate assumption of action from left to right. 

The logical structure of the proof---with the two very different statuses of (1) the input 
assumption of no action from right to left and (2) the resulting output conclusion of a necessary 
transfer of information from left to right---is revealed far more clearly and directly in the 
fixed-past-open-future formulation of the conditions for applicability of two propositions than 
in an approach that mixes counterfactual concepts into the meanings of the words appearing in 
the proofs. If that latter approach is used, then it is necessary in principle to unpack the 
counterfactual statements in order to clearly distinguish between legitimate inputs and 
possible illegitimate ones. Shimony's counterfactual-based analysis fails make this crucial 
distinction. In lieu of making this distinction within the counterfactual approach, the alternative 
and simpler way to verify the validity of the basic claim is to work directly from the assumptions 
of my 2004 paper, in the way described above, and thereby circumvent the subtlties introduced 
by the avoidable use of counterfactuals. 
 
ACKNOWLEDGEMENTS
 
I thank J. Finkelstein for very useful comments on an earlier much shorter draft of this paper. 
This work was supported in part by the Director, Office of Science, Office of High Energy and 
Nuclear Physics of the U.S. Department of Energy under contract No. DE-AC03-76SF00098.

REFERENCES

1. Abner Shimony, ``An Analysis of Stapp's `A Bell-type theorem without hidden variables' '', 
Foundations of Physics, Preceding Paper.

2. Henry P. Stapp, ``A Bell-type theorem without hidden variables'', 
American Journal of Physics, 72, 30-33 (2004).
 
3.  Henry P. Stapp,  ``Nonlocal character of quantum theory'', 
American Journal of Physics, 65, 300-304 (1997).

4.  John S. Bell, ``On the Einstein-Podolsky-Rosen paradox'', Physics 1, 195-200 (1964).

5 J.F. Clauser and A. Shimony, ``Bell's theorem: Experimental tests and implications'', 
Reports on Progress in Physics 41, 1881-1927 (1978).
    
6.  S. Tomonaga, ``On a relativistically invariant formulation of the quantum theory of wave fields,''
  Prog. Theor. Phys.1, 27-42  (1946).	
 
7. J. Schwinger, ``The theory of quantized fields I,'' Phys. Rev. 82, 914-927 (1951).

\end{document}